\documentclass[prc,twocolumn,showpacs,superscriptaddress, twoside,showpacs]{revtex4}
\usepackage{color,graphicx}
\usepackage{amsmath}
\usepackage{dcolumn}
\usepackage{CJK}                 
\usepackage{bm}                      
\usepackage{booktabs}
\usepackage{soul}
\usepackage{enumitem}
\usepackage{footmisc}

\begin{document}
\begin{CJK*}{GBK}{song}
\title{Microscopic description of triaxiality in Ru isotopes with covariant energy density functional theory}

\author{Z. Shi}
\affiliation{School of Physics and Nuclear Energy Engineering,
Beihang University, Beijing 100191, China}
\author{Z. P. Li}
\email{zpliphy@swu.edu.cn}
\affiliation{School of Physical Science and Technology,
Southwest University, Chongqing 400715, China}

\begin{abstract}
\begin{itemize}
\item[] \textbf{Background:}
The triaxiality in nuclear low-lying states has attracted
great interests for many years. Recently, the reduced transition probabilities for levels near the
ground state in $^{110}$Ru have been measured and provided strong evidences for a triaxial shape of this nucleus.
\item[] \textbf{Purpose:}
The aim of this work is to provide a microscopic study of low-lying states for
the Ru isotopes with $A\sim100$ and to examine in detail the role
of triaxiality, and the evolution of quadrupole shapes with the isospin and spin
degrees of freedom.
\item[]\textbf{Method:}
The low-lying excitation spectra and transition probabilities of even-even Ru isotopes are described at the beyond mean-field level by solving
a five-dimensional collective Hamiltonian with parameters determined by constrained self-consistent mean-field
calculations based on the relativistic energy density functional PC-PK1.
\item[]\textbf{Results:}
The calculated energy surfaces, low-energy spectra, intraband and interband transition rates, as well as some
characteristic collective observables, such as $E(4_{\rm g.s.}^+)/E(2^+_{\rm g.s.})$, $E(2^+_\gamma)/E(4^+_{\rm g.s.})$,
$B(E2; 2^+_{\rm g.s.}\to 0^+_{\rm g.s.})$, and $\gamma$ band staggerings are in a good agreement with the available
experimental data.
\item[]\textbf{Conclusions:}
The main features of the experimental low-lying excitation spectra and
electric transition rates are well reproduced, and thus strongly support the onset of triaxiality
in the low-lying excited states of the Ru isotopes around $^{110}$Ru.
\end{itemize}

\end{abstract}

\date{\today}

\pacs{21.10.Re, 21.60.Ev, 21.60.Jz, 27.60.+j}

\maketitle
\section{Introduction}
The triaxial deformation in atomic nucleus is related to many interesting phenomena, such as the wobbling motion~\cite{bohr1975nuclear} and nuclear chirality~\cite{frauendorf1997tilted,meng2006possible}. The observation of chiral doublet bands~\cite{starosta2001chiral} and wobbling bands~\cite{odegaard2001evidence} provides direct evidences for the existence of triaxial deformation. Recently, not only the static triaxial deformation, but also the triaxial shape transition along e.g., the increasing neutron numbers, have attracted a lot of attentions~\cite{sarriguren2008shape,robledo2009role,li2010microscopic,toh2013evidence,yao2014microscopic,nikvsic2014microscopic,sun2014spectroscopy,bhat2014nature,xiang2016novel,nomura2016structural}.

Due to the subtle interplay between single-particle and collective degrees of freedom, the evolution of the triaxiality in the $A\sim100$ mass region has been a hot topic for years; one of the examples is the Ruthenium isotopes. In the past decades, lots of experimental and theoretical efforts have been reported. In 1980s, the multiple Coulomb excitation experiments for $^{104}$Ru which is the heaviest stable Ru isotope have suggested a phase transition from spherical to a soft triaxial rotor rather than to an axially symmetric rotor with increasing neutron number in Ru isotopes~\cite{stachel1982riaxiality}. Later, through the $\beta$ decays of Tc isotopes, the low-lying collective structures of $^{106,108}$Ru~\cite{stachel1984collective}, and $^{110,112}$Ru~\cite{aysto1990collective} have been studied, which suggested the importance of triaxial deformation in all these nuclei and demonstrated a trend of increasing triaxial rigidity with more neutrons. Furthermore, the high-spin structures in these neutron-rich Ru isotopes have been extensively studied by the fusion-fission reactions~\cite{deloncle2000high} and the spontaneous fissions of $^{248}$Cm~\cite{shannon1994role} and $^{252}$Cf~\cite{zhu2007triaxiality,luo2009evolution}. In Ref.~\cite{luo2009evolution}, a pair of $\Delta I$ = 1 negative parity doublet bands were observed in $^{110}$Ru and $^{112}$Ru. They were interpreted as soft chiral vibrations. Very recently, a multi-step Coulomb excitation measurement following the post-acceleration of an unstable $^{110}$Ru beam was performed and the newly measured reduced transition probabilities provided a direct evidence for a relatively rigid triaxial shape near the ground state in $^{110}$Ru~\cite{doherty2017triaxiality}.

Theoretically, various methods~\cite{moller2006global,moller2008axial,sorgunlu2008triaxiality,boyukata2010description,faisal2010shape,nomura2010formulating,bhat2012microscopic,nomura2016structural,abusara2017triaxiality,zhang2015theoretical,chen2017effective} have been devoted to investigate the evolution of the triaxiality in Ru isotopes. Using the macroscopic-microscopic finite-range liquid-drop model (FRLDM), M\"{o}ller $et~al.$, have identified $^{108}$Ru as having the largest effects of triaxial deformation, $\sim0.7$ MeV, on the ground-state energy~\cite{moller2006global,moller2008axial}. In the interacting boson model (IBM), the $\gamma$-soft behaviors are found for the Ru isotopes around $A\sim100$, but no candidates for a triaxial ground state are found~\cite{sorgunlu2008triaxiality,boyukata2010description}. Within the framework of cranked shell model (CSM) with nonaxial deformed Woods-Saxon potential, the ground states of $^{108,110,112}$Ru are found to be triaxial, and $^{112}$Ru is the softest in $\gamma$ direction. It is also found that the ground state of $^{114}$Ru is oblate~\cite{faisal2010shape}. The potential energy surfaces (PESs) obtained from the Skyrme Hartree-Fock calculations show triaxial shapes for the ground states of $^{108-114}$Ru, which become more rigid with increasing neutron number~\cite{nomura2010formulating}. While the PESs obtained from the Hartree-Fock-Bogoliubov (HFB) with Gogny functional D1M predict triaxial ground state shapes for $^{104-114}$Ru, and the $\gamma$-soft behaviors are found, in which $^{104}$Ru is the softest in the $\gamma$ direction~\cite{nomura2016structural}. In Ref.~\cite{abusara2017triaxiality}, a prolate-triaxial-oblate shape transition is found for the isotopes $^{96-112}$Ru by the relativistic Hartree-Bogoliubov (RHB) calculations with DD-PC1 and DD-ME2; the ground states of $^{110,112}$Ru are oblate with clearly triaxial softness. The recent investigation of band structures in $^{108,110,112}$Ru with two complementary theoretical models, cranked Hartree-Fock-Bogoliubov (CHFB) method with density functional UNEDF0 and triaxial projected shell model (TPSM), concludes that the high-spin behavior in $^{108,110,112}$Ru consists with triaxial rotation while the obtained triaxial minima are fairly shallow~\cite{zhang2015theoretical}. Very recently, the band structures in these even Ru isotopes was also investigated in the framework of the effective field theory~\cite{chen2017effective}.

In spite of numerous efforts, it is clear that more studies are still necessary to draw an unambiguous conclusion on the detail of triaxial shape transition in neutron-rich Ru isotopes. On the other hand, the increasing data in this mass region accumulated by the modern experimental techniques could provide the stringent examination for various theoretical models.

The microscopic density functional theory (DFT), which starts from an effective nucleon-nucleon interaction and self-consistently determines the nuclear mean-field by all the independent particles inside, has achieved a lot of successes in describing the properties for both the nuclear ground states and excitation states. The covariant DFT (CDFT)~\cite{ring1996relativistic,vretenar2005relativistic,meng2006relativistic,meng2013progress,meng2016relativistic} embeds the fundamental Lorentz invariance from the very beginning and naturally includes the spin-orbit interaction~\cite{meng1999pseudospin,chen2003pseudospin,zhou2003spin,liang2015hidden}, which proves to be a successful theory used over the whole nuclide chart, from relatively light systems to superheavy nuclei~\cite{long2004new,zhang2005magic,meng2006relativistic,zhao2010new,zhang2013description}, from the valley of $\beta$ stability to the drip lines~\cite{meng1996relativistic,meng1998giant,meng2002giant,meng2006relativistic,meng2015halos}, and from collective rotations to collective vibrations~\cite{liang2008spin,liang2009isospin,zhao2011novel,zhao2011antimagnetic,zhao2012covariant,zhao2015rod,zhao2015impact,zhao2017multiple}. One of the most successful density functionals is PC-PK1~\cite{zhao2010new}, which could provide a good description for the isospin dependence of nuclear properties, such as mass~\cite{zhao2010new,zhao2012crucial}, quadrupole moments~\cite{zhao2014explanation}, etc.
\begin{figure*}[t!]
\centering
\includegraphics[scale=0.5,angle=0]{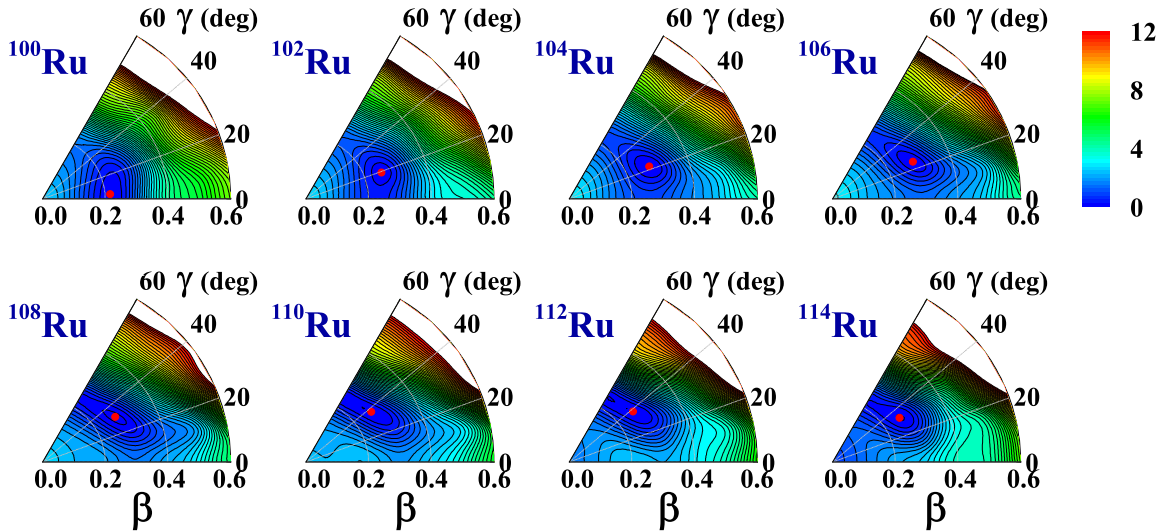}
\caption{(Color online) The potential energy surfaces of even-even $^{100-114}$Ru isotopes in the $\beta$-$\gamma$ plane calculated by the constrained triaxial RMF+BCS with PC-PK1 functional. All energies are normalized with respect to the binding energy of the absolute minimum (labeled by red dot). The energy difference between the neighbouring contour lines is 0.25 MeV.}
\label{pes-ru}
\end{figure*}

To take into account the beyond mean-field effects and describe the low-lying states, in the past few years, the five-dimensional collective Hamiltonian based on CDFT (5DCH-CDFT) has been developed~\cite{nikvsic2009beyond,li2009microscopic,nikvsic2011relativistic} and achieved great success as for nuclei ranging from light to superheavy mass regions~\cite{quan2017global}, including the spherical~\cite{li2012enhanced,wang2015covariant}, transitional~\cite{li2009microscopic02,li2010microscopic,li2011energy,xiang2012covariant,fu2013beyond,nikvsic2014microscopic,xiang2016novel,wang2016tidal}, and well-deformed~\cite{nikvsic2009beyond,li2009microscopic,li2010relativistic} ones. The approach of collective Hamiltonian has also been applied to the chiral~\cite{chen2013collective,chen2016two} and wobbling~\cite{chen2014collective,chen2016wobbling,chen2016two} motions. In the $A\sim100$ mass region, the triaxial structures in the Mo isotopes, the neighbouring even element of Ru, and the $N$ = 60 isotones have been investigated with the 5DCH-CDFT~\cite{xiang2016novel}. It is found that the evolution of nuclear collectivity is governed by novel triaxial structure that the triaxiality serves as a tunnel from the weakly deformed oblate shape to the largely deformed prolate shape~\cite{xiang2016novel}.

To provide a new survey on the shape transitions in Ru isotopes and a microscopic description on the recent experimental data, in this paper we present the 5DCH-CDFT studies to the even-even $^{100-114}$Ru isotopes based on the density functional PC-PK1. A systematic analysis that includes collective potential energy surfaces, low-energy collective spectra, electric transition rates, $\gamma$ band staggerings, and collective wave functions are carried out.


\section{Theoretical Framework}\label{sec1}
The detailed formalism of the 5DCH has been presented in a number of literatures, see, e.g. Refs.~\cite{prochniak1999collective,nikvsic2009beyond,delaroche2010structure}. For completeness, a brief introduction is presented here. The 5DCH, which could simultaneously treat the quadrupole vibrational and rotational excitations, is expressed in terms of the two deformation parameters $\beta$ and $\gamma$, and three Euler angles $(\phi,\theta,\psi)\equiv \Omega$ that define the orientation of the intrinsic principal axes in the laboratory frame,
\begin{equation}
    \hat H_{\rm coll}(\beta,\gamma)=\hat T_{\rm vib}(\beta,\gamma)
    +\hat T_{\rm rot}(\beta,\gamma,\Omega)+V_{\rm coll}(\beta,\gamma).\label{eq1}
\end{equation}
The three terms in $\hat H_{\rm coll}(\beta,\gamma)$ are the vibrational kinetic energy
\begin{align}
\hat{T}_{\rm vib}=&-\frac{\hbar^2}{2\sqrt{wr}}\left\{\frac{1}{\beta^4}\left[\frac{\partial}{\partial \beta}\sqrt{\frac{r}{w}}\beta^4B_{\gamma\gamma}\frac{\partial}{\partial \beta}
-\frac{\partial}{\partial\beta}\sqrt{\frac{r}{w}}\beta^3 B_{\beta\gamma}\frac{\partial}{\partial\gamma}\right] \right.\\ \notag
&+\frac{1}{\beta \sin 3\gamma}\left[-\frac{\partial}{\partial\gamma}
\sqrt{\frac{r}{w}}\sin3\gamma B_{\beta\gamma}\frac{\partial}{\partial\beta} \right. \\ \notag
&\left.\left.+\frac{1}{\beta}\frac{\partial}{\partial\gamma}\sqrt{\frac{r}{w}}
\sin 3\gamma B_{\beta\beta}\frac{\partial}{\partial\gamma}\right]\right\},
\label{eq2}
\end{align}
the rotational kinetic energy
\begin{equation}
   \hat T_{\rm rot}=\frac{1}{2}\sum^3_{k=1}\frac{\hat J_k^2}{\mathcal I_k},
   \label{eq3}
\end{equation}
and the collective potential $V_{\rm coll}$, respectively. Here, $\hat J_k$ denote the components of the total angular momentum in the body-fixed frame, and both the mass parameters $B_{\beta\beta}$, $B_{\beta\gamma}$, $B_{\gamma\gamma}$ and the moments of inertia $\mathcal I_k$ depend on the quadrupole deformation variables $\beta$ and $\gamma$. Two additional quantities that appear in the $\hat{T}_{\rm vib}$ term, $r = B_1B_2B_3$ and $w = B_{\beta\beta}B_{\gamma\gamma} - B_{\beta\gamma}^2$, determine the volume element in the collective space.

The eigenvalue problem of the Hamiltonian~(\ref{eq1}) is solved using an expansion of eigenfunctions  in terms of a complete set of basis functions that depend on the five collective coordinates  $\beta, \gamma$ and $\Omega~(\phi,\theta,\psi)$~\cite{nikvsic2009beyond}. Using the collective  wave functions thus obtained
\begin{align}
    \Psi^{IM}_\alpha(\beta, \gamma, \Omega)=\sum_{K\in\Delta I}\psi^I_{\alpha K}(\beta, \gamma)\Phi^I_{MK}(\Omega),
\end{align}
various observables such as the $E2$ transition probabilities can be calculated,
\begin{align}
    B(E2;\alpha I\rightarrow\alpha'I')=\frac{1}{2I+1}|\langle\alpha'I'||\hat M(E2)||\alpha I\rangle|^2,
\end{align}
where $\hat M (E2)$ is the electric quadrupole operator.

In the framework of 5DCH-CDFT, the collective parameters of 5DCH, including the mass parameters $B_{\beta\beta}$, $B_{\beta\gamma}$, and  $B_{\gamma\gamma}$, the moments of inertia $\mathcal I_k$, and the collective potential $V_{\rm coll}$, are all determined microscopically from the constrained triaxial CDFT calculations. The moments of inertia are calculated with Inglis-Belyaev formula~\cite{inglis1956nuclear,beliaev1961concerning} and the mass parameters with the cranking approximation~\cite{girod1979zero}. The collective potential $V_{\rm coll}$ is obtained by subtracting the zero-point energy corrections~\cite{girod1979zero} from the total energy that corresponds to the solution of constrained triaxial CDFT. Detailed formalism can be found in Ref.~\cite{nikvsic2009beyond}.

\section{Results and Discussion}\label{sec2}
In the present work, we are focusing on the even Ru isotopes with the neutron number from $N=54$ to 70.
To determine the collective parameters in the 5DCH, we perform the constrained triaxial CDFT calculations; the pairing correlations are treated with the BCS method. In the particle-hole channel, the point-coupling density functional PC-PK1~\cite{zhao2010new} is used, and a density-independent $\delta$-force is used in the particle-particle channel. The strength parameter of the $\delta$-force is 349.5 MeV fm$^3$ (330.0 MeV fm$^3$) for neutrons (protons)~\cite{zhao2010new}. The Dirac equation is solved by expanding the Dirac  spinor in terms of the three dimensional harmonic oscillator basis with 12 major shells.

\subsection{Potential energy surfaces and binding energies}

Figure~\ref{pes-ru} displays the PESs of even-even $^{100-114}$Ru isotopes in the $\beta$-$\gamma$ plane. The quadrupole deformations $(\beta$, $\gamma)$ that correspond to the global minima are listed in Table~\ref{global-minimum-ru}, so are the energy differences of $\Delta E_{\rm tri}$. In the following, we denote this energy difference as the triaxial deformation energy. The $\Delta E_{\rm tri}$ is defined as the energy difference of the global minimum with respect to the lowest energy under axial symmetry. Starting from the nearly prolate $^{100}$Ru, where the global minimum locates at $(0.22,4^\circ)$, a considerable triaxial deformation, $\gamma=19^\circ$, is predicted in  the global minimum of $^{102}$Ru with only two more neutrons. The PES of $^{102}$Ru is rather soft along $\gamma$ direction with $|\Delta E_{\rm tri}|=0.324$ MeV. The patterns of the PESs in $^{104, 106}$Ru are similar with remarkable triaxiality, $\gamma>20^\circ$ and  $|\Delta E_{\rm tri}|>0.6$ MeV. For $N\geq64$, the deformations of the global minima of $^{108}$Ru, $^{110}$Ru, $^{112}$Ru, and $^{114}$Ru are $(0.27,32^\circ)$, $(0.26, 37^\circ)$, $(0.26,38^\circ)$, and $(0.25, 34^\circ)$, respectively. The PESs of these four nuclei are very flat along $\gamma$ direction towards the oblate side but relatively rigid towards the prolate one. It is also noted that a local spherical minimum emerges in $^{114}$Ru.

{
\begin{table}[t!]\footnotesize
\centering
\caption{The quadrupole deformations $(\beta, \gamma)$ of the global minima and the triaxial deformation energies $\Delta E_{\rm tri}$ for $^{100-114}$Ru calculated by the CDFT with PC-PK1.
$E_{\rm tri}$ is the total energy for the global minima, and $E_{\rm axi}$ is the lowest energy under axial symmetry.
}
\begin{tabular}{ccccc}
\hline
\hline
  Nucleus~~& $(\beta, \gamma)$&~~$E_{\rm tri}$ (MeV)&~~$E_{\rm axi}$ (MeV)&~~$\Delta E_{\rm tri}$ (MeV) \\
\hline
$^{100}$Ru~~&$(0.21, 4^\circ)$ &~~ -858.425 &~~ -858.421 &~~ -0.004  \\
$^{102}$Ru~~&$(0.25, 19^\circ)$&~~ -874.342 &~~ -874.018 &~~ -0.324  \\
$^{104}$Ru~~&$(0.27, 22^\circ)$&~~ -889.607 &~~ -888.739 &~~ -0.868  \\
$^{106}$Ru~~&$(0.28, 25^\circ)$&~~ -904.040 &~~ -903.364 &~~ -0.676  \\
$^{108}$Ru~~&$(0.27, 32^\circ)$&~~ -917.718 &~~ -917.408 &~~ -0.310  \\
$^{110}$Ru~~&$(0.26, 37^\circ)$&~~ -930.674 &~~ -930.601 &~~ -0.073  \\
$^{112}$Ru~~&$(0.26, 38^\circ)$&~~ -942.767 &~~ -942.712 &~~ -0.055  \\
$^{114}$Ru~~&$(0.25, 33^\circ)$&~~ -953.971 &~~ -953.611 &~~ -0.360  \\
\hline
\hline
\end{tabular}\label{global-minimum-ru}
\end{table}
}

\begin{figure}[htbp]
\centering
\includegraphics[scale=0.33,angle=0]{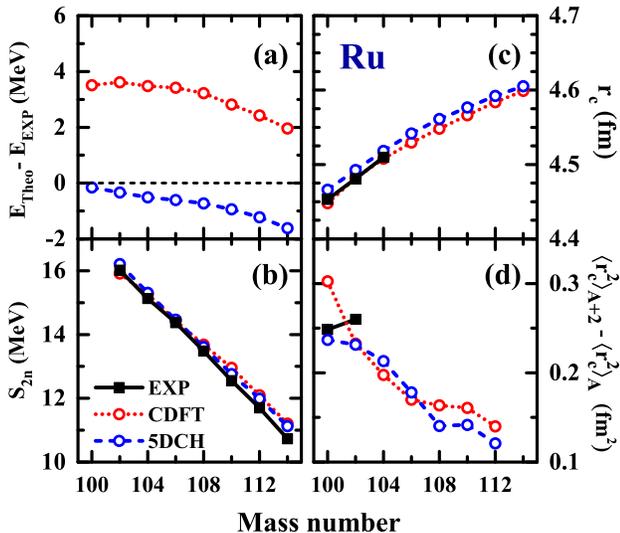}
\caption{(Color online) (a) Deviations of the binding energies calculated by the triaxial CDFT and 5DCH from the data for Ru isotopes. Evolution of the theoretical two-neutron separation energies $S_{2n}$ (b),
  root-mean-square charge radii $r_c$ (c), and isotope shifts of the ground-state charge radii
  $\langle r^2_c\rangle_{A+2}-\langle r^2_c\rangle_A$ (d) as functions of mass number in Ru isotopes, in
  comparison with available data \cite{wang2017ame2016,angeli2013table}.}
\label{mass-ru}
\end{figure}

\begin{figure*}
\centering \hspace{-2.3cm}
  \begin{minipage}[t]{0.3\textwidth}
  \includegraphics[scale=0.32]{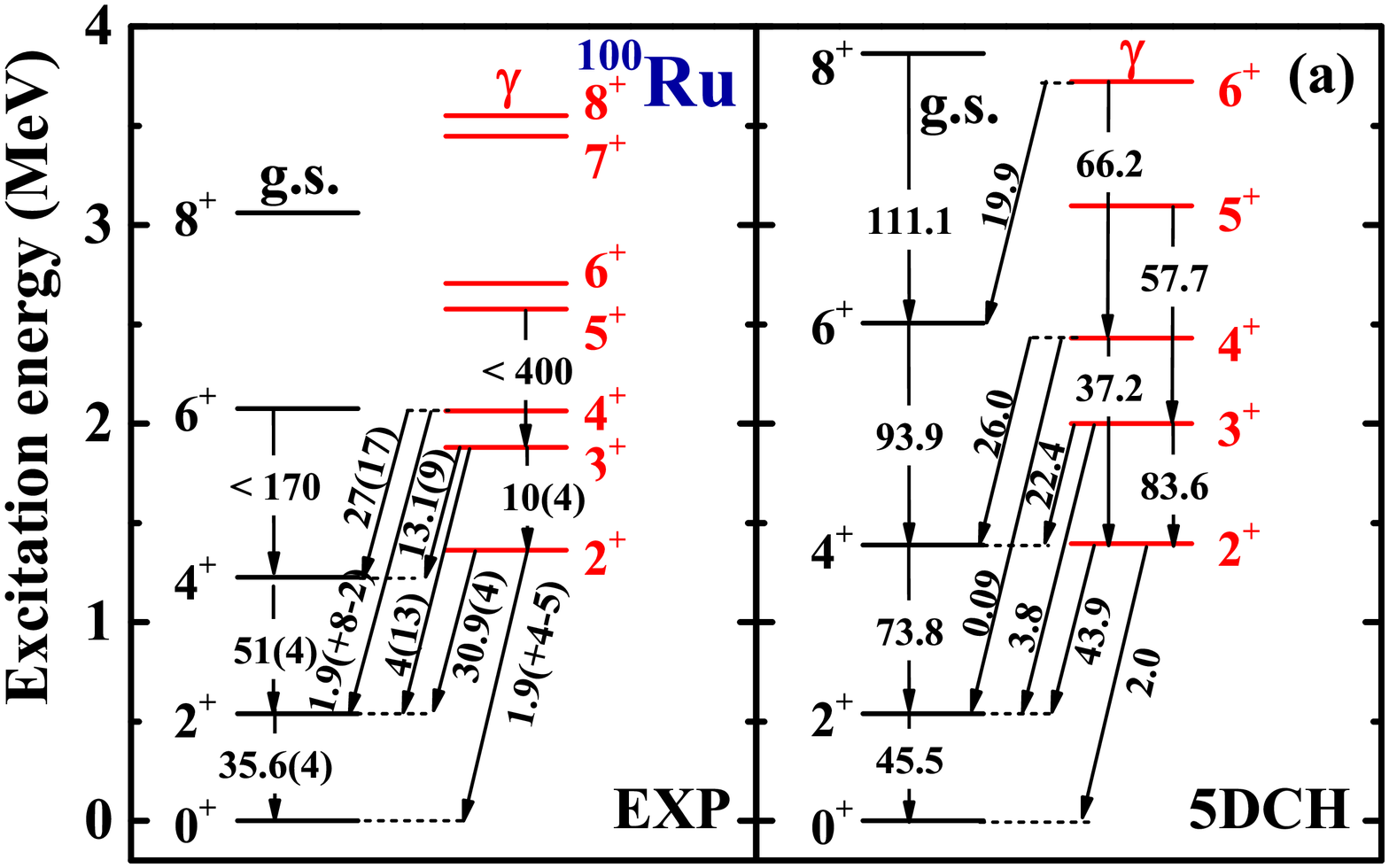}\vspace{0.2cm}
  \includegraphics[scale=0.32]{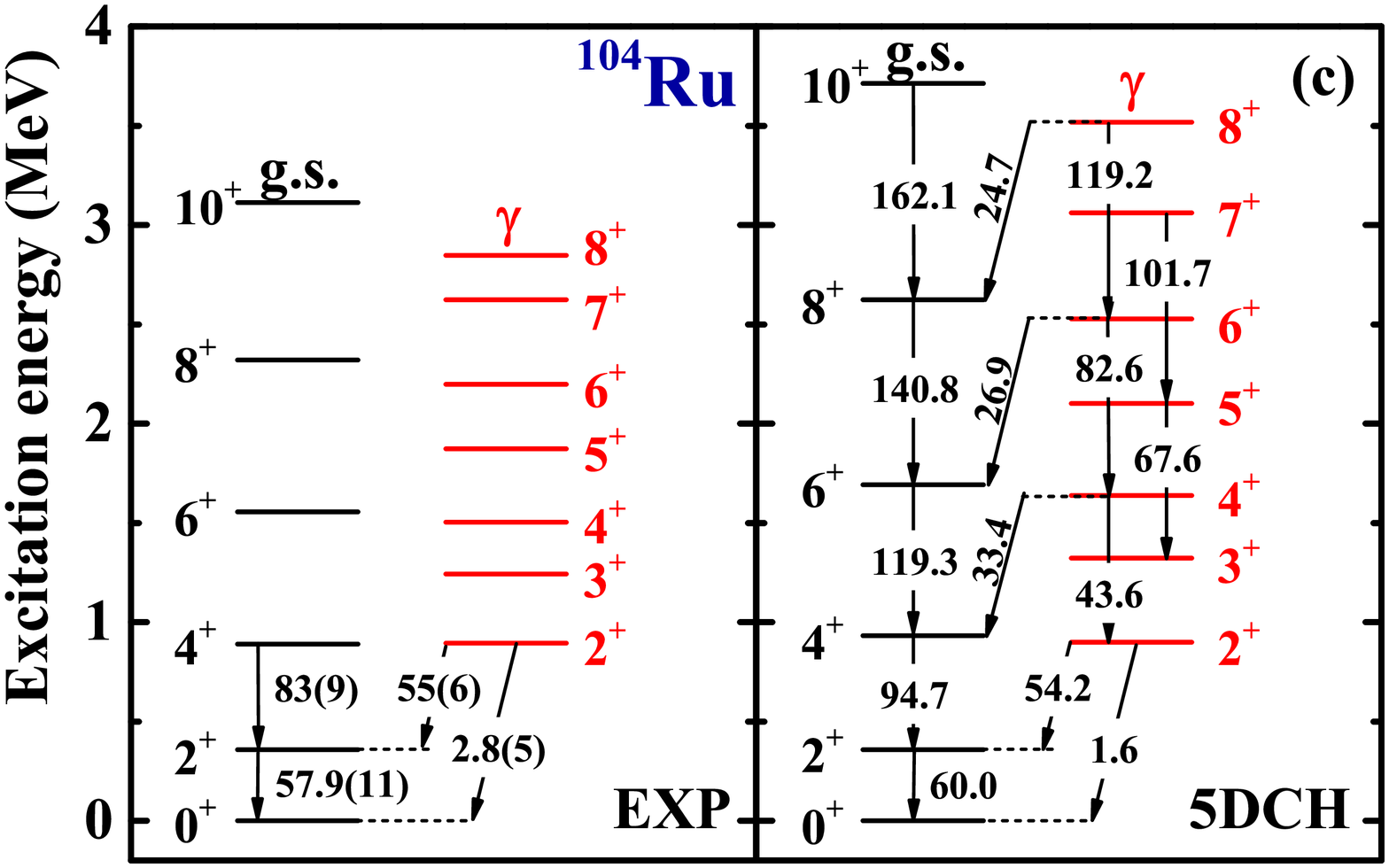}\vspace{0.2cm}
  \includegraphics[scale=0.32]{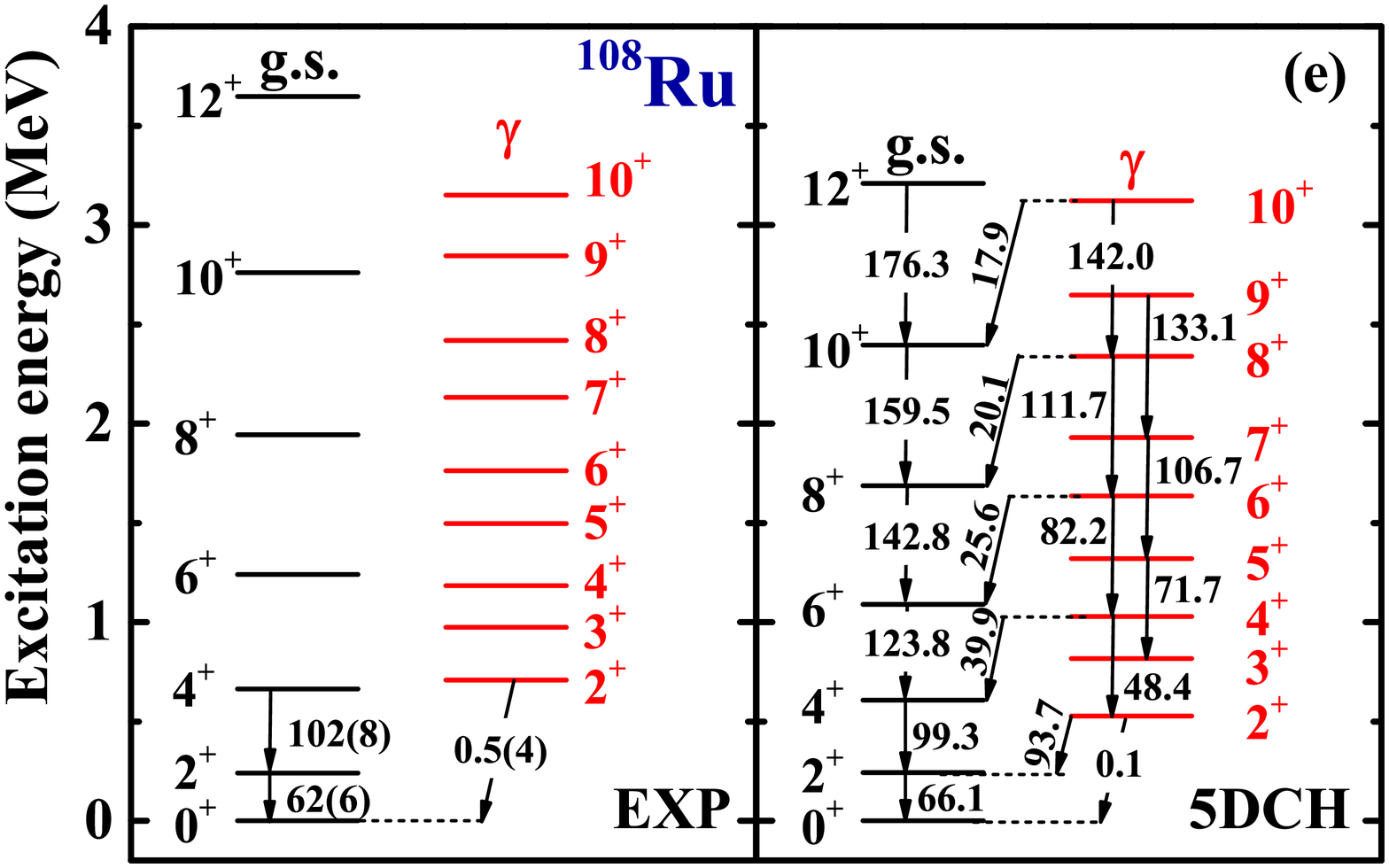}\vspace{0.2cm}
  \includegraphics[scale=0.32]{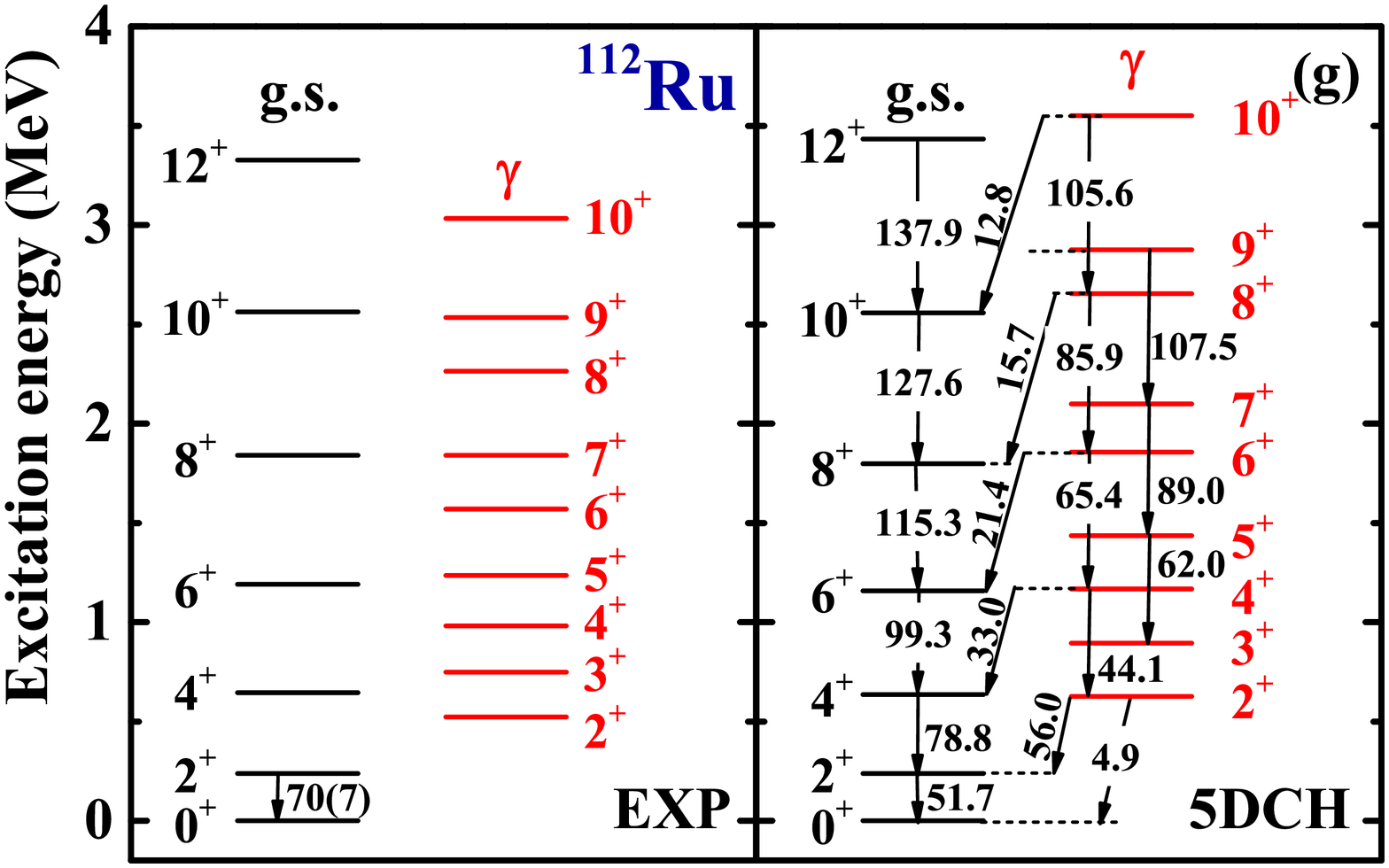}\vspace{0.2cm}
  \end{minipage}
  \hspace{3cm}
  \begin{minipage}[t]{0.3\textwidth}
  \includegraphics[scale=0.32]{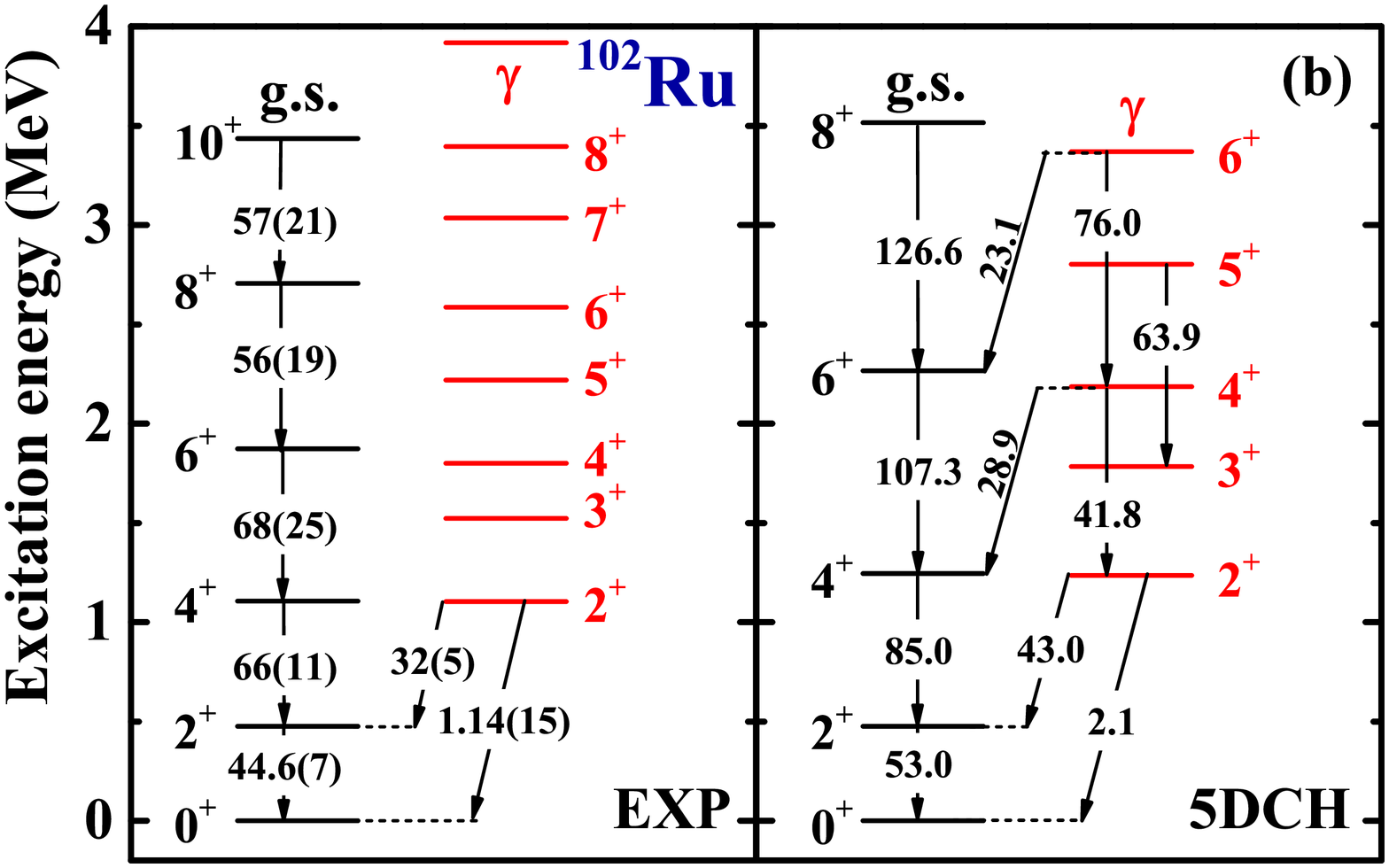}\vspace{0.2cm}
  \includegraphics[scale=0.32]{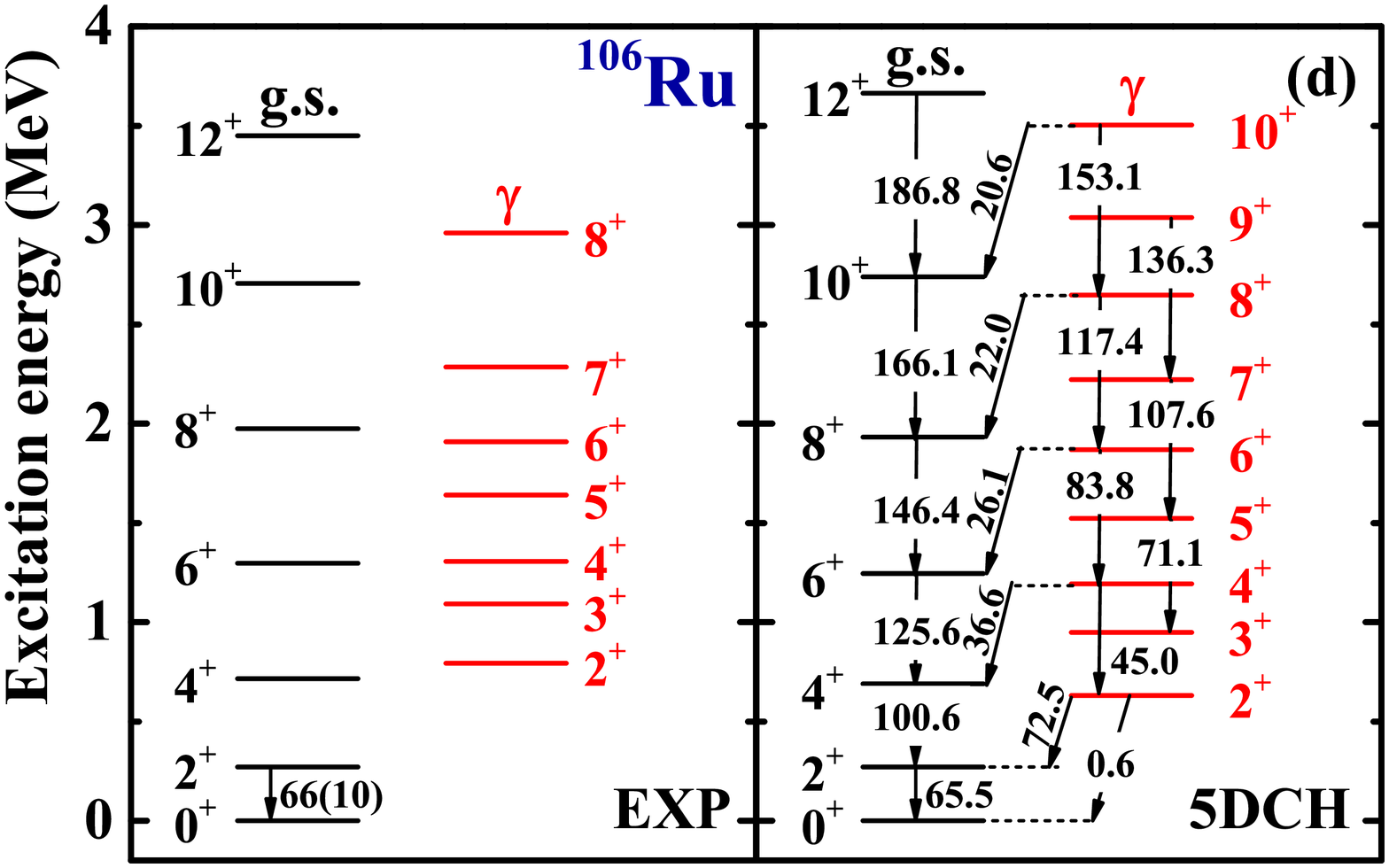}\vspace{0.2cm}
  \includegraphics[scale=0.32]{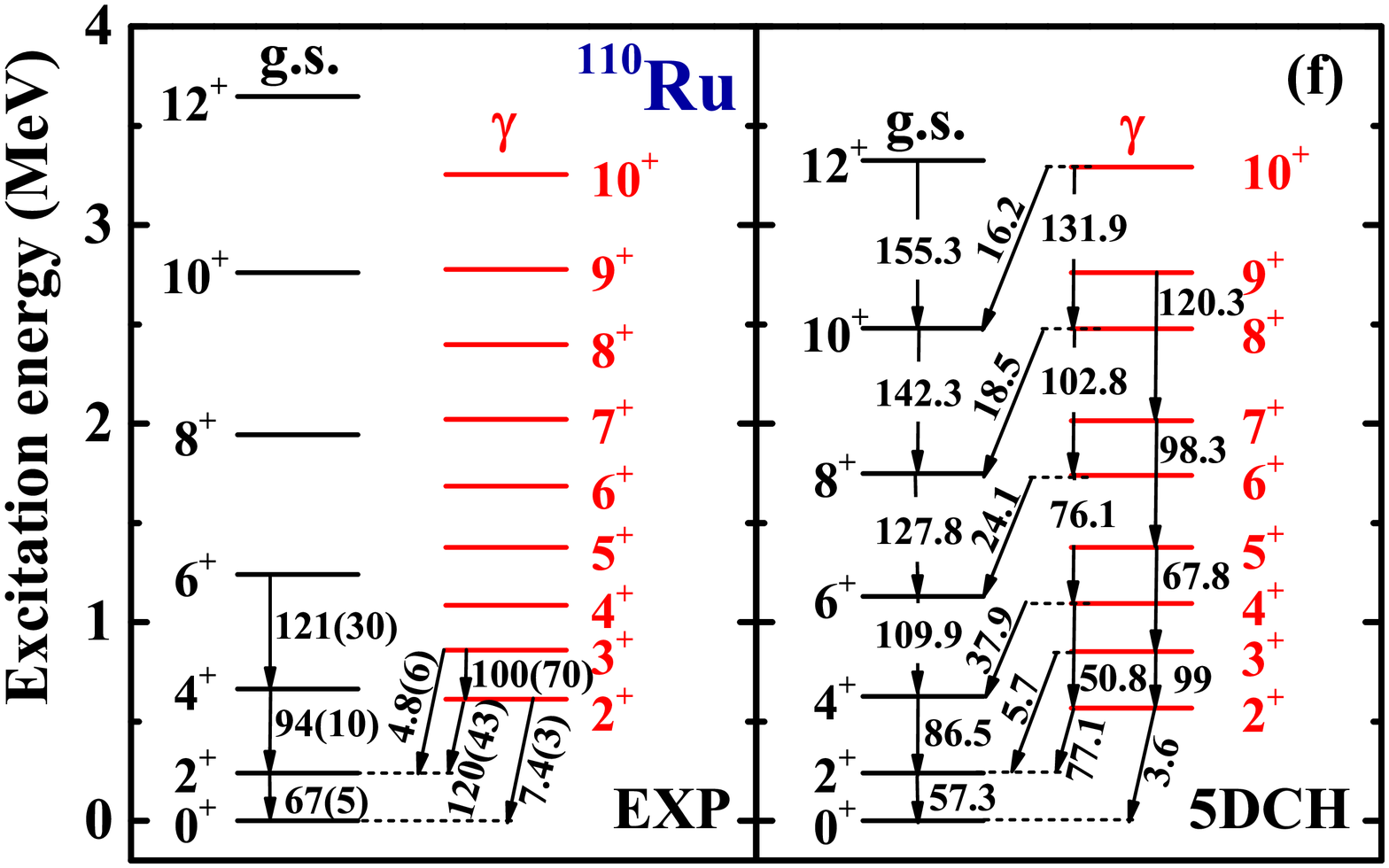}\vspace{0.2cm}
  \includegraphics[scale=0.32]{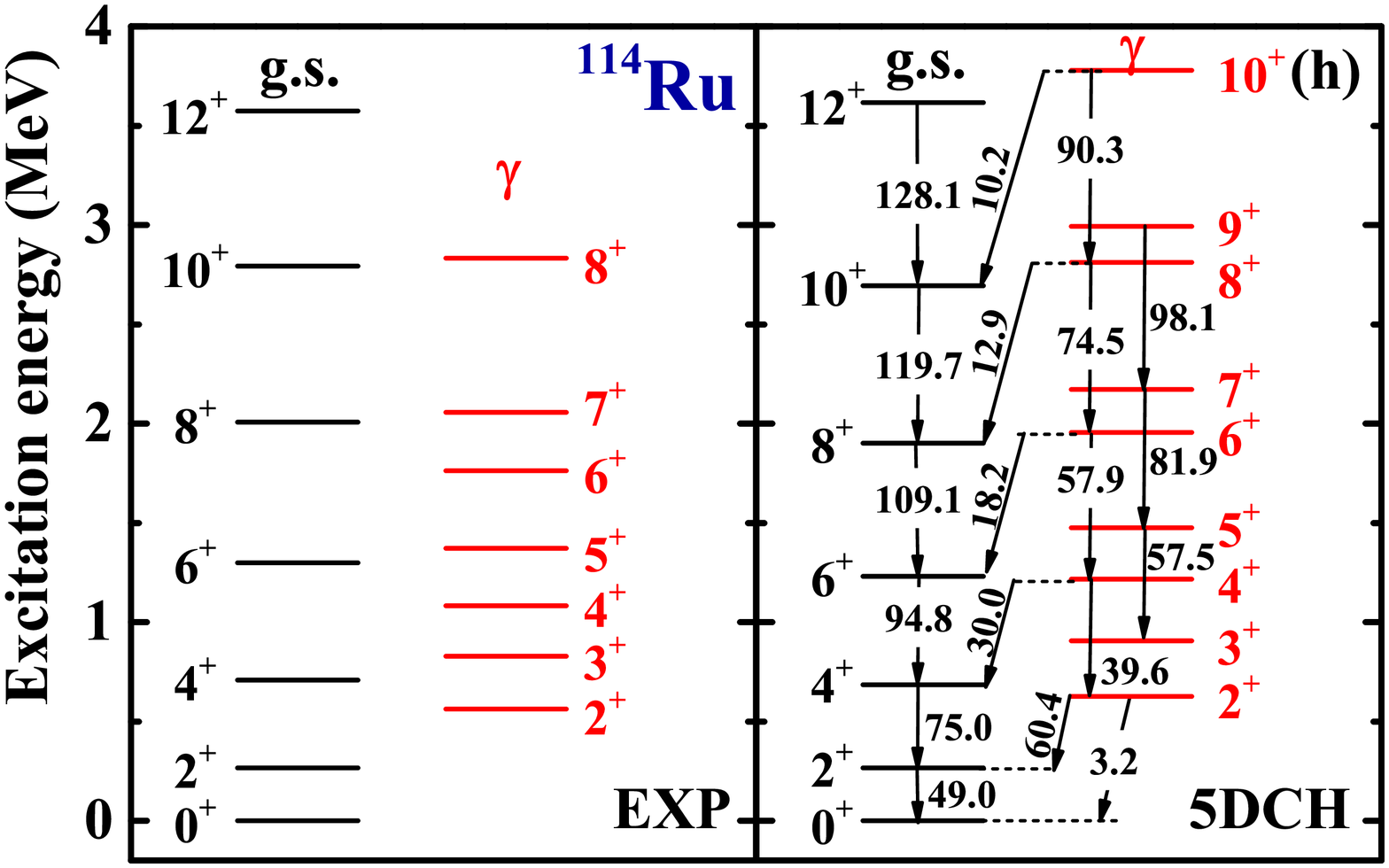}\vspace{0.2cm}
  \end{minipage}
  \caption{(Color online) The calculated excitation energies (in MeV) and the intraband and interband $B(E2)$ values (in W.u.) for the ground-state bands and $\gamma$  bands in even-even $^{100-114}$Ru isotopes by 5DCH-CDFT, in comparison with the experimental data\protect\footnote{In the present work, the data  for $^{104}$Ru are taken from the evaluated database in National Nuclear Data Center (NNDC) \cite{http://www.nndc.bnl.gov/}, while in Ref.~\cite{xiang2016novel} we used the data from [http: //ie.lbl.gov/TOI2003/index.asp.].}~\cite{http://www.nndc.bnl.gov/,doherty2017triaxiality}.}
  \label{spectra-ru}
\end{figure*}

\begin{figure*}[t!]
  \centering
  \includegraphics[scale=0.4,angle=0]{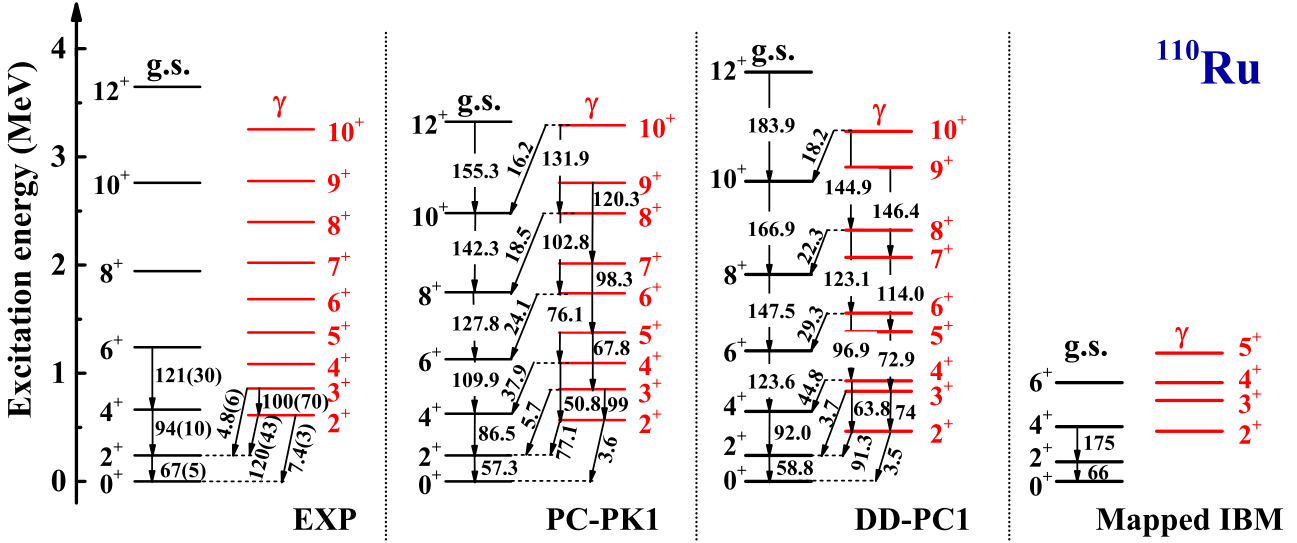}
  \caption{(Color online) The energy spectra for $^{110}$Ru calculated by 5DCH-CDFT with PC-PK1~\cite{zhao2010new}
  and DD-PC1~\cite{niksic2008relativistic}, in comparison with the available mapped IBM results~\cite{nomura2016structural}
  and experimental data~\cite{http://www.nndc.bnl.gov/,doherty2017triaxiality}. In the calculation with DD-PC1, a separable pairing~\cite{tian2009finite} is adopted.}
\label{ru110-compare}
\end{figure*}

\begin{figure}[h!]
\centering
\includegraphics[scale=0.45,angle=0]{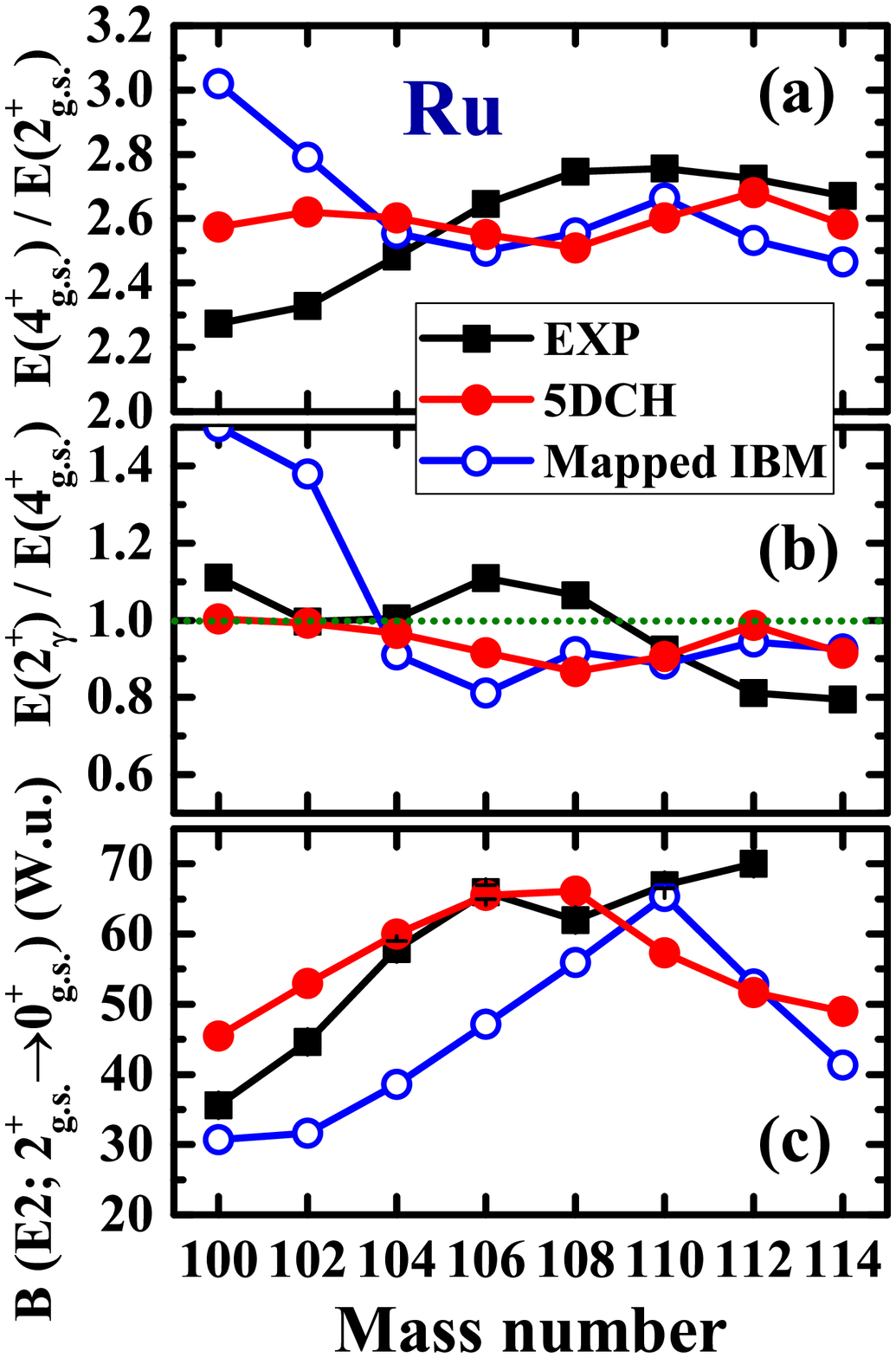}
\caption{(Color online) Evolution of $E(4_{\rm g.s.}^+)/E(2^+_{\rm g.s.})$ (a), $E(2^+_\gamma)/E(4^+_{\rm g.s.})$ (b), and $B(E2; 2^+_{\rm g.s.}\to 0^+_{\rm g.s.})$ values (in W.u.) with mass number for Ru isotopes calculated by 5DCH-CDFT, in comparison with the available data~\cite{http://www.nndc.bnl.gov/,doherty2017triaxiality} and theoretical results from mapped IBM~\cite{nomura2016structural}.}
\label{r42-e2}
\end{figure}

As mentioned above, similar topograghy of the PESs in Ru isotopes have also been obtained in the studies based on the RHB with density functionals DD-ME2 and DD-PC1~\cite{abusara2017triaxiality}, and the HFB with Gogny-D1M density functional~\cite{nomura2016structural}. Some differences can be found in the exact locations of the equilibrium triaxial minima and the corresponding triaxial deformation energies. The ground states of $^{110,112}$Ru are oblate in the RHB calculations with both DD-ME2 and DD-PC1 functionals~\cite{abusara2017triaxiality}. In the HFB calculation with Gogny-D1M, only $^{104-108}$Ru are triaxial deformed at ground states. The shape coexistence in $^{114}$Ru is also not observed in the HFB calculation~\cite{nomura2016structural}.

In Figs.~\ref{mass-ru}(a) and \ref{mass-ru}(b), we compare the theoretical binding energies and two-neutron separation energies calculated by the triaxial CDFT and the 5DCH to the experimental data for Ru isotopes. For the mean-field calculations, the deviations of the binding energies from the data are in $2.0\sim3.5$ MeV. It is remarkable that the deviations are reduced to be within 1.6 MeV by considering the dynamical correlations associated with rotational motion and quadrupole shape vibrational motion in the 5DCH calculations. This is consistent with a global study of dynamic correlation energies (DCEs) for 575 even-even nuclei by using the 5DCH based on the PC-PK1 functional~\cite{zhang2014global,lu2015global}. In the global study in Refs.~\cite{zhang2014global,lu2015global}, after taking into account these DCEs, the root-mean-square deviation of the nuclear masses is reduced significantly from 2.52 to 1.14 MeV~\cite{lu2015global}. The description of two-neutron separation energies is slightly modified by the 5DCH; both the mean-field and 5DCH results are in good agreement with the data.

In Figs.~\ref{mass-ru}(c) and \ref{mass-ru}(d), the charge radii $r_c$ and isotope shifts of the ground-state charge radii $\langle r^2_c\rangle_{A+2}-\langle r^2_c\rangle_A$ as functions of mass number in Ru isotopes are shown.
The charge radii calculated by CDFT are in good agreement with available data and increase smoothly with mass number. Similar results are obtained by 5DCH, but slightly larger than those of CDFT because of the beyond-mean-field effect \cite{delaroche2010structure}. The theoretical isotope shifts decrease gradually as the mass number increases, that is the increasing trend of charge radii becomes slow because of the saturation of quadrupole deformations (c.f. Fig.~\ref{pes-ru} and Table \ref{global-minimum-ru}).
\subsection{Low-energy collective spectra}
Starting from constrained self-consistent solutions, the collective parameters that determine 5DCH are calculated as functions of the deformation parameters $\beta$ and $\gamma$. The diagonalization of the resulting Hamiltonian yields the excitation energies and collective wave functions.

Figure~\ref{spectra-ru} displays the collective excitation spectra, including the ground-state bands and $\gamma$ bands, in the even-even $^{100-114}$Ru isotopes calculated by the 5DCH-CDFT, in comparison with the available experimental data. The intraband and interband $B(E2)$ values are also shown in the figure. It is noted that the inertia parameters in the present study are calculated by the Inglis-Byleav formula, which do not include the Thouless-Valatin dynamical rearrangement contributions and, thus would systematically underestimate the empirical values. As illustrated in Ref.~\cite{li2012efficient}, the Thouless-Valatin corrections are almost independent on deformation and, therefore, for a given nucleus the effective moment of inertia used in the collective Hamiltonian can simply be obtained by renormalizing the Inglis-Byleav values with a  constant factor, which is determined by reproducing the excitation energy of the $2^+_1$ state~\cite{nikvsic2009beyond}.

\begin{figure*}[htbp]
\centering
\includegraphics[scale=0.4,angle=0]{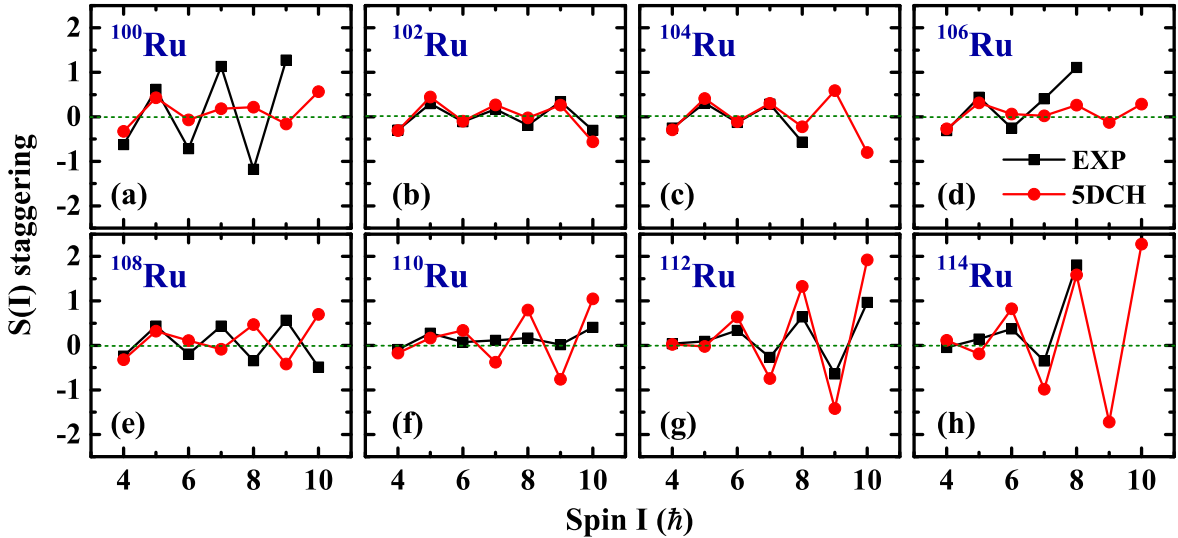}
\caption{(Color online) The staggering parameters $S(I)$ of even-even $^{100-114}$Ru isotopes calculated by 5DCH-CDFT in comparison with the available data~\cite{http://www.nndc.bnl.gov/}.}
\label{staggering}
\end{figure*}

The levels are grouped into ground-state bands (labeled as g.s.) and $\gamma$ bands (labeled as $\gamma$) according to the predominant $K$ components and dominant decay patterns. For the stable nuclei $^{100,102}$Ru, the 5DCH calculations can reproduce the collective structure although the theoretical spectra are stretched and the intraband transitions are generally larger. This may be due to the overestimation of the collectivity of these two isotopes in the calculations. Starting from $^{104}$Ru, the 5DCH calculations are in very good agreement with the experimental data for both excitation energies and transition rates. In particular, the signatures of the triaxiality including the low-lying $\gamma$ bandhead, the enhanced interband transitions between the $\gamma$ band and ground-state band, the $\gamma$ band staggerings (c.f. Fig. \ref{staggering}), and the relations $E(3^+_\gamma)\approx E(2^+_{\rm g.s.})+E(2^+_\gamma)$~\cite{davydov1958rotational}, are all reproduced very well.

Very recently, the reduced transition probabilities obtained for levels near the ground state of $^{110}$Ru have been measured and provided strong evidences for a triaxial shape of this nucleus~\cite{doherty2017triaxiality}. As seen in Fig.~\ref{ru110-compare}, the measured intraband and interband $B(E2)$ values of $^{110}$Ru are consistent with our 5DCH calculations with both PC-PK1~\cite{zhao2010new} and DD-PC1~\cite{niksic2008relativistic}
functionals, and also the mapped IBM calculation~\cite{nomura2016structural}. It should be emphasized that in the 5DCH model, the transition probabilities are calculated in a full configuration space and there are no effective charges used. Therefore, the agreements between the present calculations and experimental data are very remarkable. Moreover, the excitation energies predicted by both PC-PK1 and DD-PC1 functionals are also in very good agreement with the data. Combining the calculated PES, low-energy spectrum, and $E2$ transition rates of $^{110}$Ru, the triaxiality near the ground state of $^{110}$Ru is further supported.

Furthermore, in Fig.~\ref{r42-e2} we analyze the evolution of some characteristic collective observables, such as $E(4_{\rm g.s.}^+)/E(2^+_{\rm g.s.})$, $E(2^+_\gamma)/E(4^+_{\rm g.s.})$, and $B(E2; 2^+_{\rm g.s.}\to 0^+_{\rm g.s.})$ values with mass number of Ru isotopes calculated by the 5DCH-CDFT, in comparison with the available data~\cite{http://www.nndc.bnl.gov/} and theoretical results from mapped IBM~\cite{nomura2016structural}. The measured $E(4_{\rm g.s.}^+)/E(2^+_{\rm g.s.})$ values vary from $\sim2.3$ to $\sim2.7$, indicating that the Ru isotopes locate in a transitional region. However, for $^{100, 102}$Ru, the theoretical results calculated by both the 5DCH and the mapped IBM are too large. This is probably because both calculations overestimate the collectivity of these nuclei. For heavier isotopes, the experimental values of $E(4_{\rm g.s.}^+)/E(2^+_{\rm g.s.})$ are reproduced by the 5DCH and mapped IBM quite well.

For a nucleus with considerable triaxial deformation, the $\gamma$ bandhead $2^+_\gamma$ is generally lower than the state of $4^+_{\rm g.s.}$, and this is fulfilled for $^{110-114}$Ru according to the measurement in Fig.~\ref{r42-e2} (b). The calculated values of $E(2^+_\gamma)/E(4^+_{\rm g.s.})$ by the 5DCH are in a reasonable agreement with the data, and the possible triaxial deformation is predicted to start from $^{106}$Ru. This is similar to the calculations from the mapped IBM, but for $^{100,102}$Ru, the mapped IBM overestimates the data significantly. This is due to the fact that the IBM model space, comprising only a finite number of $s$ and $d$ bosons, is not large enough to describe the energy levels near the closed shell~\cite{nomura2016structural}. The comparison of $B(E2; 2_{\rm g.s.}^+\rightarrow0_{\rm g.s.}^+)$ among the 5DCH, the mapped IBM~\cite{nomura2016structural} and the experimental data are shown in Fig.~\ref{r42-e2} (c). The experimental $B(E2; 2_{\rm g.s.}^+\rightarrow0_{\rm g.s.}^+)$  increases gradually till $A=106$ and saturates for heavier isotopes. Without any effective charges, the 5DCH can reproduce the experimental data very well, except for $A\geq110$, where the theoretical results decrease with mass number. On the other hand, the results from the mapped IBM are overall smaller than the data.

The $\gamma$ band staggering parameter
\begin{equation}
 S(I)=\frac{\left[E(I)-E(I-1)\right]-\left[E(I-1)-E(I-2)\right]}{E(2^+_1)} \nonumber
\end{equation}
is an indicator of the triaxial softness/rigidness~\cite{zamfir1991signatures}. For a nucleus with a deformed $\gamma$-soft potential, $S(I)$ oscillates between negative values for even-spin states and positive values for odd-spin states, with the magnitude slowly increasing with spin. For a triaxial potential, the level clustering in the $\gamma$ band is opposite, and $S(I)$ oscillates between positive values for even-spin states and negative values for odd-spin states. In this case, the magnitude of $S(I)$ increases more rapidly with spin, as compared to the $\gamma$-soft potential~\cite{McCutchan2007staggering}.

\begin{figure*}[t!]
\centering
\includegraphics[scale=0.7,angle=0]{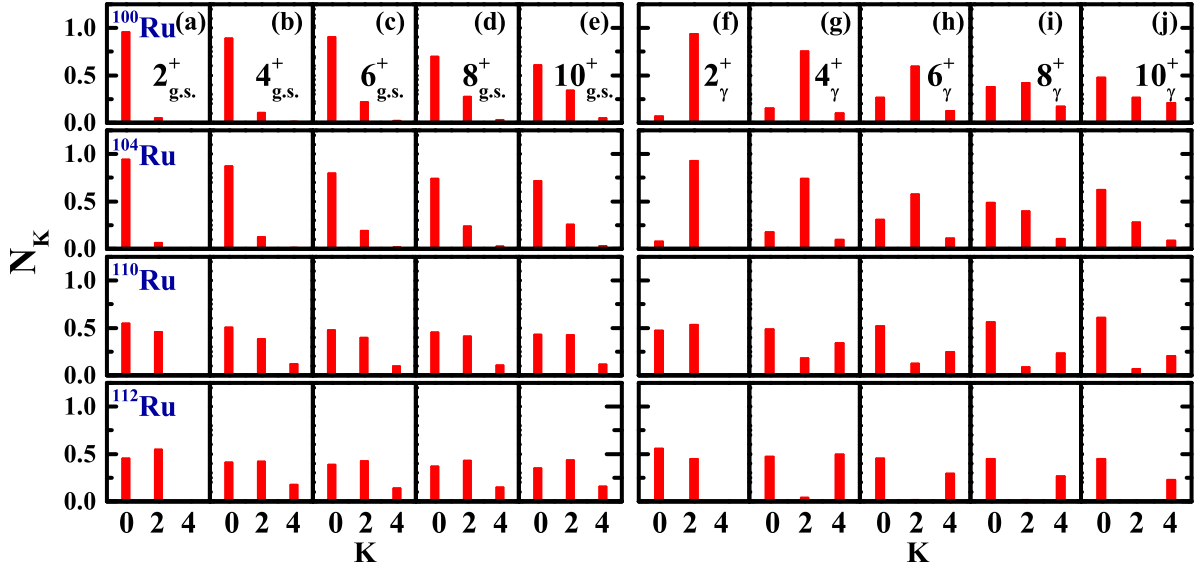}
\caption{(Color online) Distributions of $K$ components in the collective wave functions for the $2^+$, $4^+$, $6^+$, $8^+$, and $10^+$ states in the ground-state and $\gamma$ bands of $^{100,104, 110, 112}$Ru.}
\label{k-ground-ru}
\end{figure*}

In Fig.~\ref{staggering}, we plot the theoretical $\gamma$ band staggering parameters $S(I)$ for Ru isotopes, in comparison with the available experimental data. In general, the experimental staggering parameters are well reproduced by the 5DCH calculations, in particular for low spins. For the isotopes $^{100-104}$Ru, the $\gamma$ band staggering parameters $S(I)$ present as the cases with deformed $\gamma$-soft potentials, namely oscillating between negative values for even-spin states and positive values for odd-spin states. The deviation for the high spin states in $^{100}$Ru may be because the calculated PES is too
stiff in the $\gamma$ direction around the global minimum (c.f. Fig.~\ref{pes-ru}). Moving to $^{106}$Ru, the phase of $S(I)$ at low spins is same to the case of $\gamma$-soft potential but inverses for $I\geq8$ $\hbar$. Therefore, it could be a transitional nucleus from $\gamma$-soft to triaxial deformed shape along increasing isospin. The nucleus $^{108}$Ru has a $\gamma$-soft potential according to the experimental $S(I)$, while the calculated $S(I)$ demonstrates that this nucleus is similar to the neighboring $^{106}$Ru as a transitional nucleus. The model probably overestimates the triaxiality of $^{108}$Ru, which is also reflected in
the too low $\gamma$ bandhead $2^+_\gamma$ and too large interband $B(E2; 2^+_\gamma\to 2^+_{g.s.})$ in Fig.~\ref{spectra-ru}(e). For $^{110}$Ru, although the $S(4)$ is negative and $S(5)$ is positive, both of them are very close to zero.  When $I\geq6\hbar$, the $S(I)$ becomes negative for odd-spins and positive for even-spins, and considerable oscillation amplitudes are also observed. Thus, $^{110}$Ru is close to the case of $\gamma$-rigid shape~\cite{davydov1958rotational}. Remarkable oscillations of $S(I)$ are observed in $^{112,114}$Ru, indicating that they are triaxiality deformed.

\begin{figure}[htbp]
\centering
\includegraphics[scale=0.4,angle=0]{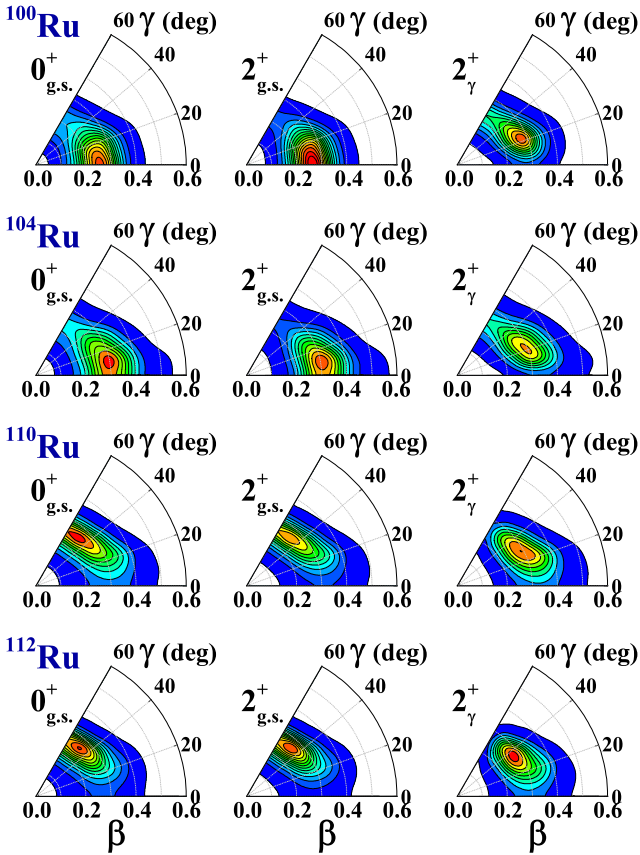}
\caption{(Color online) Probability density distributions in $\beta$-$\gamma$ plane for $0_{\rm g.s.}^+$, $2_{\rm g.s.}^+$ and $2_\gamma^+$ states in $^{100, 104, 110, 112}$Ru.}
\label{wf-ru}
\end{figure}

The mixing of different intrinsic configurations in the state $|\alpha I\rangle$ can be demonstrated from the distribution of $K$, the projection of angular momentum $I$ on the third axis in the body-fixed frame:
\begin{align}
    N_K=6\int^{\pi/3}_0\int^\infty_0|\psi^I_{\alpha K}(\beta,\gamma)|^2\beta^4|\sin3\gamma|d\beta d\gamma.
\end{align}
Figure~\ref{k-ground-ru} displays the distributions of $K$ components in the collective wave functions for the $2^+$, $4^+$, $6^+$, $8^+$, and $10^+$ states in the ground-state and $\gamma$ bands of selected Ru isotopes: $^{100}$Ru, $^{104}$Ru, $^{110}$Ru, and $^{112}$Ru. In the cases of $^{100, 104}$Ru, the $K=0$ components are predominant in the wave functions for the yrast states, whereas the states comprising $\gamma$ bands are dominated by the $K = 2$ components. The mixing of $K=0$ and $K=2$ components is small for low-spin states, and thus the states are dominated by the prolate configures (c.f. Fig.~\ref{wf-ru}). For the higher spin states, the mixing of $K$ components is much stronger, indicating that the triaxial degree of freedom plays an important role in these states. For $^{110, 112}$Ru, the mixing of $K$ components becomes remarkable for all the states and it is also notable that the $K>2$ components are pronounced in the $\gamma$ bands. This is strongly correlated to the triaxial deformed potentials of these isotopes (c.f. Fig.~\ref{pes-ru}) and also consistent with the oscillation behavior of $S(I)$ in Fig.~\ref{staggering}.

The density distribution of the collective state, which takes the following form,
\begin{align}
    \rho_{I\alpha}(\beta,\gamma)=\sum_{K\in\Delta I}|\psi^I_{\alpha K}(\beta,\gamma)|^2\beta^3,
\end{align}
with the normalization
\begin{align}
    \int^\infty_0\beta d\beta\int^{2\pi}_0\rho_{I\alpha}(\beta,\gamma)|\sin(3\gamma)|d\gamma=1,
\end{align}
could give a further insight into the shape evolution with spin and isospin. Here, taking $^{100}$Ru, $^{104}$Ru, $^{110}$Ru, and $^{112}$Ru as examples, the density distributions for the $0_{\rm g.s.}^+$, $2_{\rm g.s.}^+$ and $2_\gamma^+$ states are depicted in Fig.~\ref{wf-ru}. For the states $0^+$ and $2^+$ in the ground-state bands of $^{100, 104, 110, 112}$Ru, the peaks of collective wave functions are in general consistent with the global minima of the PESs, as shown in Fig.~\ref{pes-ru}, while somewhat differences are observed because the masses of inertia are strongly deformation dependent. Weak triaxial deformations are predicted in the states $0_{\rm g.s.}^+$ and $2_{\rm g.s.}^+$ of $^{104, 110, 112}$Ru isotopes. The collective wave functions of $2_\gamma^+$ all concentrate on the region with $\gamma=20^\circ\sim40^\circ$, which demonstrates the importance of the triaxial degree of freedom in this mass region of Ru isotopes.

\section{Summary}\label{sec3}
In this work, we have presented a microscopic and systematic beyond mean-field investigation for the low-lying states in the Ru isotopes around $A\sim100$ mass region. The excitation energies and transition strengths calculated from a five-dimensional collective Hamiltonian with parameters determined from the constrained triaxial CDFT calculations with PC-PK1 functional reproduce the available data well. The microscopic potential energy surfaces exhibit transitions with increasing neutron number: from prolate $^{100}$Ru to triaxial $^{114}$Ru. The low-energy spectra, interband transition rates between the $\gamma$ band and the ground-state band, collective wave functions, as well as characteristic collective observables $E(4^+_{\rm g.s.})/E(2^+_{\rm g.s.})$, $E(2^+_\gamma/E(4^+_{\rm g.s.})$, $B(E2; 2^+_{\rm g.s.}\to 0^+_{\rm g.s.})$, and $\gamma$ band staggerings strongly support the onset of triaxiality in low-lying states of Ru isotopes around $^{110}$Ru.

\begin{acknowledgments}
ZS is indebted to Prof. Jie Meng for constructive guidance and valuable suggestions. Fruitful discussions and critical readings from Dr. Shuangquan Zhang and Dr. Pengwei Zhao are highly acknowledged. This work was partly supported by the National Natural Science Foundation of China (Grants No. 11475140, No. 11375015, No. 11461141002, No. 11335002, No. 11621131001), the Chinese Major State 973 Program (Grant No. 2013CB834400).
\end{acknowledgments}


\end{CJK*}
\end{document}